\documentclass[twocolumn]{aastex631}
\usepackage{graphicx}

\begin{document}

\title{The Nature of the IMBH Candidate CXO~J133815.6+043255: High-Frequency Radio Emission}

\author[0000-0001-5785-7038]{Krista Lynne Smith}
\affiliation{George P. and Cynthia Woods Mitchell Institute for Fundamental Physics and Astronomy, Texas A\&M University, College Station, TX 77843-4242, USA}
\affiliation{Southern Methodist University, Department of Physics
Dallas, TX 75205}

\author{Macon Magno}
\affiliation{George P. and Cynthia Woods Mitchell Institute for Fundamental Physics and Astronomy, Texas A\&M University, College Station, TX 77843-4242, USA}
\affiliation{Southern Methodist University, Department of Physics
Dallas, TX 75205}
\affiliation{CSIRO Space and Astronomy, ATNF, PO Box 1130, Bentley WA 6102, Australia}

\author{Ashutosh Tripathi}
\affiliation{George P. and Cynthia Woods Mitchell Institute for Fundamental Physics and Astronomy, Texas A\&M University, College Station, TX 77843-4242, USA}
\affiliation{Southern Methodist University, Department of Physics
Dallas, TX 75205}



\begin{abstract}

The ultra-luminous X-ray source CXO~J133815.6+043255 is a strong candidate for a bona-fide intermediate mass black hole, residing in the outskirts of NGC~5252. We present 22~GHz radio observations of this source obtained serendipitously in an ongoing high-frequency imaging survey of radio-quiet Active Galactic Nuclei (AGN), and use this new data point to construct the broad-band radio spectral energy distribution (SED). We find that the SED exhibits a spectral slope of $\alpha=-0.66\pm0.02$, consistent with a steep spectrum from optically-thin synchrotron emission from an unresolved jet. We also find that the $L_R / L_X$ ratio is approximately $10^{-3}$, inconsistent with radio-quiet AGN and many ULXs but consistent with low-luminosity AGN (LLAGN) and radio-loud quasars. Together, these observations support the conclusion that CXO~J133815.6+043255 is an intermediate-mass black hole producing a low-mass analog of radio jets seen in classical quasars.
\end{abstract}



\section{Introduction} \label{sec:intro}

Intermediate mass black holes (IMBHs) fall in the mass gap between stellar remnants ($M_\mathrm{BH}\sim10 M_\odot$) and their supermassive counterparts  ($M_\mathrm{BH}\sim10^6-10^{10} M_\odot$). Evidence for black holes in the lower portion of this intermediate range is now commonplace in gravitational wave detections from the Laser Interferometer Gravitational Wave Observatory (LIGO), which routinely observes mergers of $\sim10 M_\odot$ black holes that presumably produce a remnant with $M_\mathrm{BH}\sim10^2 M_\odot$ \citep[e.g.,][]{Abbott2020}. Electromagnetic observations have long suggested the presence of black holes in this mass range, although never unambiguously. \citet{Colbert1999} first interpreted bright X-ray sources in spiral galaxies as possible black holes with masses of $10^2-10^4~M_\odot$. In the past two decades,observational focus on ultra-luminous X-ray sources (ULXs) has  generated a number of IMBH candidates. X-ray spectroscopy of ULXs has been found to be consistent with the cooler accretion disks of smaller black holes, although sometimes requiring the assumption of unusual accretion states \citep{Miller2003,Bachetti2013,Palit2023}. Measurements of the radio and X-ray luminosities have allowed mass estimates via the black hole fundamental plane \citep{Merloni2003} that suggest intermediate values for some objects \citep{Mezcua2011,Cseh2012}. The object HLX-1, located in the galaxy ESO~243-49, has an X-ray luminosity implying a strong lower bound of a few hundred solar masses that is supported by numerous electromagnetic follow-up studies \citep{Farrell2010}. X-ray timing has also delivered several candidates, from observations of quasi-periodic oscillations in several ULXs that indicate intermediate masses if typical frequency-mass scaling relations are followed \citep[e.g., M~82~X-1, NGC~5408~X-1][]{Pasham2014, Strohmayer2007}. While these objects remain viable IMBH candidates, it has become clear that the majority of ULXs are most likely to be particular accretion states of X-ray binaries, especially neutron stars accreting at many times the Eddington rate, as described in recent reviews \citep{Fabrika2021,King2023}.

One especially interesting IMBH candidate is CXO~J133815.6+043255 (hereafter CXO~J1338+04), an ultra-luminous X-ray source (ULX) in the outskirts of the lenticular Seyfert galaxy NGC~5252. This object has been carefully studied by a handful of detailed investigations over the past decade. It was discovered by \citet{Kim2015} in a search of archival \emph{Chandra} images, and confirmed to have AGN-like optical ionization line ratios by the same study. In follow up investigations, \citet{Kim2017} found evidence that the ionized gas in the vicinity was kinematically bound to the object and has very low metallicity, leading the authors to posit that the ULX was once the central black hole in a dwarf galaxy that is now in the late stages of a minor merger with NGC~5252. \citet{Mezcua2018} provided the strongest evidence of an IMBH in the source so far: a VLBI image showing a resolved, 2-component radio source at the location of the ULX, with an SED analysis suggesting that one source is the radio core, and the other a small jet lobe. Using the black hole fundamental plane \citep{Merloni2003}, they constrain the black hole mass to $10^{3.5} M_\odot < M_\mathrm{BH} < 10^{6.3} M_\odot$. Finally, \citet{Kim2020} conducted near-infrared (NIR) imaging of the source with the \emph{Herschel} space telescope to study the stellar
properties of the putative remnant dwarf host, finding a stellar mass of $M_* =10^{7.9} M_\odot$ for the remnant galaxy, and used scaling relations to estimate the IMBH mass at $10^5 M_\odot$.

\begin{figure*}[htp]
  \centering
\includegraphics[width=\textwidth]{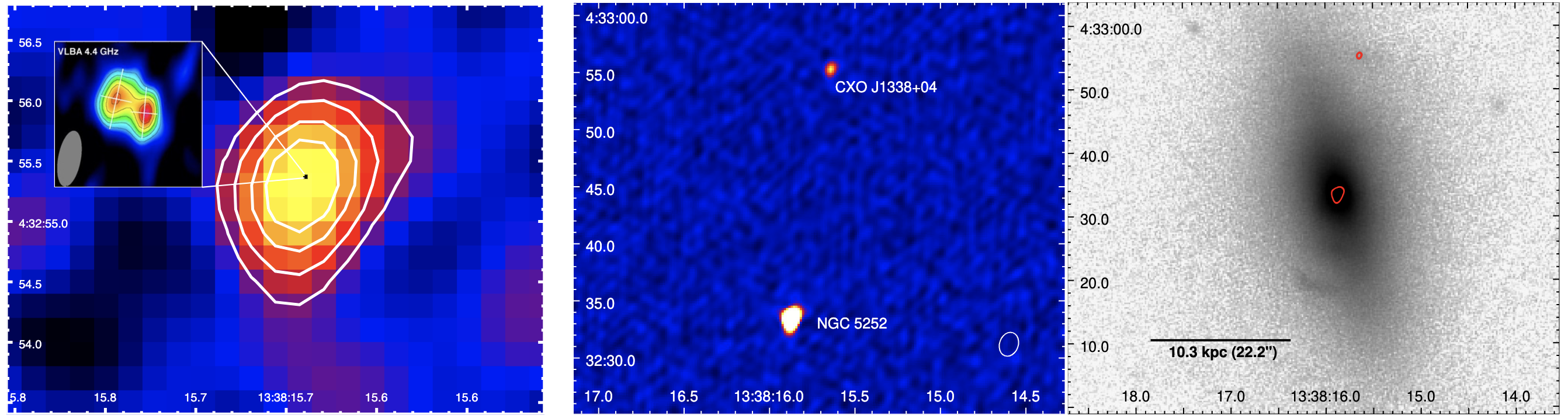}
  \caption{(a) Close-up view of our K-band radio image of ULX CXO~J1338+04. Contours occur at 70, 50, 30, and 15\% of the peak flux density. Inset is a reproduction of the VLBI image from \citet{Mezcua2018}; the very small black box in the center, from which the inset is enlarged, has the dimensions of the VLBI image. (b) \emph{Left:} Full K-band radio image of NGC~5252, including the ULX in the top right. The beam is shown on bottom right. \emph{Right: } The PanSTARRS g-band image of NGC~5252 with radio contours overlaid in red. The scalebar shows the 10.3~kpc distance between the Seyfert nucleus and the ULX.}
  \label{fig:images}
\end{figure*}

In this work, we present data obtained serendipitously for CXO~J1338+04 in our 22~GHz radio survey of nearby radio-quiet ultra-hard X-ray selected AGN from the \emph{Swift}-BAT survey conducted over the past several years with the Jansky Very Large Array (JVLA) \citep{Smith2016,Smith2020b}. 

The main target in our radio observation was NGC~5252, but our field of view easily encompasses CXO~J1338+04, which is robustly detected ($>5\sigma$). We combine our new, high-frequency radio flux density with archival radio data, and analyze the source's X-ray to radio luminosity ratio in the context of other accreting sources to place further constraints on its nature.

In Section~\ref{sec:observation}, we discuss the JVLA observation and data reduction. In Section~\ref{sec:results}, we present the broadband radio spectral energy distribution and the $L_R/L_X$ results. Section~\ref{sec:discussion} provides some discussion of the results, and a conclusive summary can be found in Section~\ref{sec:conclusions}. 

Throughout, we assume cosmological parameters $H_0 = 69.6$~km~s$^{-1}$~Mpc$^{-1}$, $\Omega_M = 0.286$, and $\Omega_\Lambda=0.714$ \citep{Bennett2014}.

\section{Observation and Data Reduction}
\label{sec:observation}
The object CXO~J1338+04 was observed in the same integration as its host galaxy NGC~5252, as part of our JVLA 22~GHz imaging survey of radio-quiet AGN host galaxies from the \emph{Swift}-BAT ultra-hard X-ray survey \citep{Baumgartner2013}. The observations were taken in the K-band while the array was in C-configuration, so the resulting images have a beam size of 1\arcsec. We have presented the results of the first two phases of this survey in three papers \citep{Smith2016,Smith2020b,Smith2020a}; it is the radio segment of the larger BAT AGN Spectroscopic Survey \citep[BASS; ][]{Koss2017,Koss2022}, a multi-wavelength follow-up effort for the full BAT AGN sample.

Our observations of NGC~5252 occurred on March 2, 2020 as part of an execution block including two other science targets. The block began with X- and K-band attenuation scans and an X-band reference pointing scan on 3C~286. The science observation of NGC~5252 had an on-source integration time of 8 minutes and 50 seconds, and was bracketed on either side by gain calibration scans of J1347+1217 and preceded by an X-band pointing calibration scan of this same target. The raw data were passed through the standard JVLA reduction
pipeline at the National Radio Astronomy Observatory (NRAO). We then processed them using the Common Astronomy Software Applications package \citep[v. 4.5, ][]{McMullin2007}, and inspected the data for radio-frequency interference (RFI) or bad phase calibration interactively with CASA's \texttt{plotms}. In the case of NGC~5252, neither effect was found. Finally, each
science target was split from the main measurement set and
averaged over all 64 channels within each spectral window, and the data were cleaned to a 0.03 mJy threshold with Briggs weighting.

The image obtained by this observation is shown in Figure~\ref{fig:images} with an inset of the VLBI image of \citet{Mezcua2018} and alongside the $g$-band image of NGC~5252 from the Pan-STARRS survey \citep{Chambers2016}. The ULX is approximately 10~kpc from the galaxy center. The radio source is unresolved in our image with $\theta_\mathrm{maj} = 1.07$\arcsec~ and $\theta_\mathrm{min} = 0.82$\arcsec; the resolution of our observations is much larger than the VLBI image from \citet{Mezcua2018} in which the pc-scale jet was resolved.

\section{Results}
\label{sec:results}
\subsection{Radio Spectral Energy Distribution}
\label{sec:radiosed}

We measure an unresolved flux density of $0.53\pm{0.02}$~mJy for CXO~J1338+04. (The Seyfert nucleus in NGC~5252 was detected at $6.39\pm{0.04}$~mJy; this and the core flux densities for our full sample will be published in the upcoming final paper in our survey series, Magno et al. in preparation.)

\begin{figure}[t]
\centering
\includegraphics[width=0.5\textwidth]{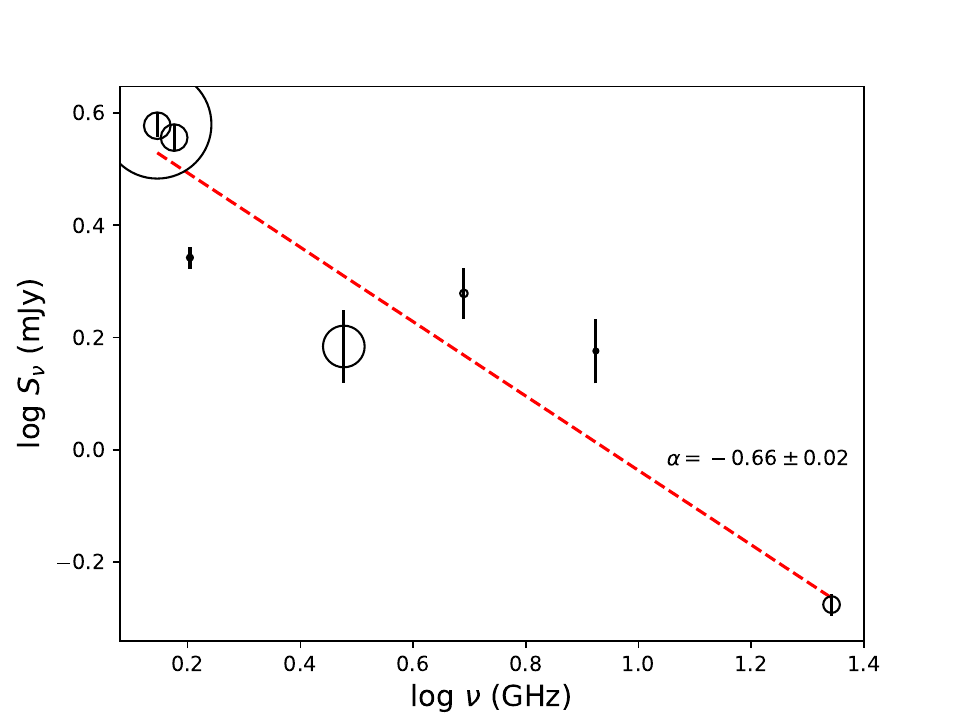}
\caption{Spectral energy distribution of CXO~J1338+04, including data summarized by \citet{Yang2017}, and our data point at 22~GHz (log~$\frac{\nu}{\mathrm{GHz}} = 1.34$). Circles represent the relative sizes of the beam for each observation; the smallest is the 0.36\arcsec~ 1.6~GHz MERLIN observations by \citet{Thean2001}, and $\sim0.3$\arcsec~ resolutions of the averaged 8.4~GHz observations by \citet{Wilson1994}, \citet{Kukula1995}, and \citet{Nagar1999}. The largest is the 1.4~GHz flux observation from the FIRST survey, at  $\sim6$\arcsec~ \citep{Becker1995}.}
\label{fig:seds}
\end{figure}

Several observations at lower radio frequencies exist in the literature. These are helpfully tabulated by \citet{Yang2017} in their Table~2. We note that they noticed no significant variability in the radio flux density despite these archival measurements occurring over several decades. Our beam size is slightly smaller than the majority of the VLA beam sizes for observations taken at 1.49~GHz ($\sim1.5$\arcsec) and slightly larger than VLA beam sizes for observations taken at 8.4~GHz ($\sim0.3\arcsec$). We construct an SED with these archival observations in Figure~\ref{fig:seds}. The flux densities of 8.4~GHz observations, taken annually from 1990 through 1993 \citep{Wilson1994,Kukula1995,Nagar1999}, are all consistent with one another and at similar resolution ($\sim0.3$\arcsec, taken in A and A/B hybrid configurations of the VLA), so we average these measurements in the plot and fitting. 

CXO~J1338+04 is also robustly detected in the VLA Sky Survey \citep[VLASS; ][]{Lacy2020} at 3~GHz and 2.5\arcsec~resolution. We used an Epoch 1.2 VLASS image centered on NGC 5252. This image was generated using the Canadian Initiative for Radio Astronomy Data Analysis (CIRADA) cutout service\footnote{\url{cutouts.cirada.ca}}. Flux density was measured using the Cube Analysis and Rendering Tool for Astronomy \citep[CARTA; ][]{Comrie2021} interface: we create a 2.5\arcsec$\times$2.0\arcsec~ region with a position angle of 18.45 degrees centered on the ULX, fully containing its emission. CARTA calculates the flux density by multiplying the mean intensity in this region by the number of beam areas subtended (where the beam area is defined as the volume of the elliptical Gaussian defined by the synthesized beam, divided by the maximum of that function). We obtain a VLASS 3~GHz flux density of $1.53\pm0.16$~mJy; however, the currently available VLASS images have a systematic error of 15\% according to its documentation 
 (0.22 on our measured flux). We use this value in the plot and the spectral slope fit.

There have also been a number of VLBI investigations of the source. \citet{Yang2017} observed it with the EVN at 1.66~GHz and find it unresolved with a flux density of $1.8\pm0.1$~mJy.
The VLBA observations of \citet{Mezcua2018} resolved the radio source into two components along a roughly east-west axis, with a combined total flux density of $0.66\pm0.09$~mJy. The eastern component is detected at both of their observing frequencies (4.4~GHz and 7.6~GHz), with a very steep spectral index between the two sources of $\alpha = -2.0 \pm 0.1$, where $S_\nu \propto \nu^\alpha$. Due to this well-constrained, steep spectral index, \citet{Mezcua2018} believe the eastern component to be a jet lobe, associated with the radio core in the western component. The western component is detected at $0.12\pm0.03$~mJy at 7.6~GHz, and is not detected at 4.4~GHz. Based on the $5\sigma$ upper limit of the flux density, they place an upper limit on the spectral index of this component of $\alpha\sim-0.6$; this flatter index is most consistent with the western source harboring the true radio core. The \citet{Yang2017} observation likely failed to resolve the two components because of the unfortunate coincidence that its beam is highly elongated along the apparent jet axis. Because of the dramatically higher resolution of these VLBI observations, we do not include them in the SED fitting, although they are relevant to the upcoming $L_R/L_X$ discussion in Section~\ref{sec:lrlx}.

Our 22~GHz data point extends the lower-resolution SED significantly in frequency space. Using least-squares minimization, we calculate an overall spectral index of $\alpha=-0.66\pm0.02$ between 1.4 and 22~GHz, a typical steep spectral index consistent with the majority of emission at these resolutions coming from an unresolved jet source.

It is clear from the VLBI-scale observations of \citet{Mezcua2018} that a jet-like radio source is present at the position of the ULX within our beam. However, since the spatial resolution of our 22~GHz observation and many of the archival observations shown in Figure~\ref{fig:seds} are not on VLBI scales, it is possible that other sources such as star-forming regions are contributing to the emission. Star-forming HII regions tend to have $\alpha \sim -0.1$ \citep{Condon1992} due to significant contributions from bremsstrahlung radiation. Isolated supernova remnants tend to have $\alpha < -0.6$ \citep[e.g., ][]{Kapinska2017}, although the relatively flat thermal component of nearby HII regions typically leads to a flatter composite spectrum of the star forming region as a whole. X-ray binaries often have flat or inverted spectra \citep[e.g., ][]{Migliari2003}. It is therefore unlikely, but not impossible, that a spectral index of $\alpha\sim-0.6$ is consistent with star formation; however, \citet{Kim2015} found AGN-like optical emission line ratios from long-slit spectroscopic observations of CXO~1338+04 with $\sim1\arcsec$ resolution, inconsistent with ratios expected from HII regions. With all of this accounted for, it is unlikely that star formation related sources are causing the steep radio spectral index reported here.

\subsection{The $L_R/L_X$ Ratio}
\label{sec:lrlx}

\citet{Laor2008} found that the relationship between the radio and X-ray luminosities of radio-quiet Seyferts and quasars is the same as that exhibited by cool, coronally active stars: $L_R/L_X \sim 10^{-5}$. In that work, the authors compare radio-loud and radio-quiet optically-selected Palomar-Green quasar subsamples \citep[PG; ][]{Schmidt1983} studied in detail by \citet{Boroson1992} and four ULXs: NGC~5408 X-1 \citep{Kaaret2003}, M82 X-1 \citep{Kaaret2006}, NGC~7424 ULX-2 \citep{Soria2006}, and Holmberg~II \citep{Dewangan2004}. As described in detail in \citet{Laor2008}, the radio fluxes of the PG sample were measured at 5~GHz by the VLA \citep{Kellermann1989,Kellermann1994} and the X-ray fluxes come from ROSAT observations by \citet{Brandt2000} and \citet{Laor2002}.  The bolometric X-ray luminosity is derived over the range 0.2-20~keV using the relation $L_X = C\nu L_\nu (1\mathrm{keV})$, with $C=6.25$; for the detailed derivation, see Section 2.1 of \cite{Laor2008}. The radio and X-ray luminosities of the four ULXs come from diverse sources as detailed in Section~2.3 of \citet{Laor2008} and the references above, but are typically derived at 5~GHz and 2-10~keV. The radio-loud and radio-quiet quasars occupy distinctly different populations in $L_R/L_X$ space, with the four ULXs consistent with the radio-quiet population.

\begin{figure*}
\centering
\includegraphics[width=0.95\textwidth]{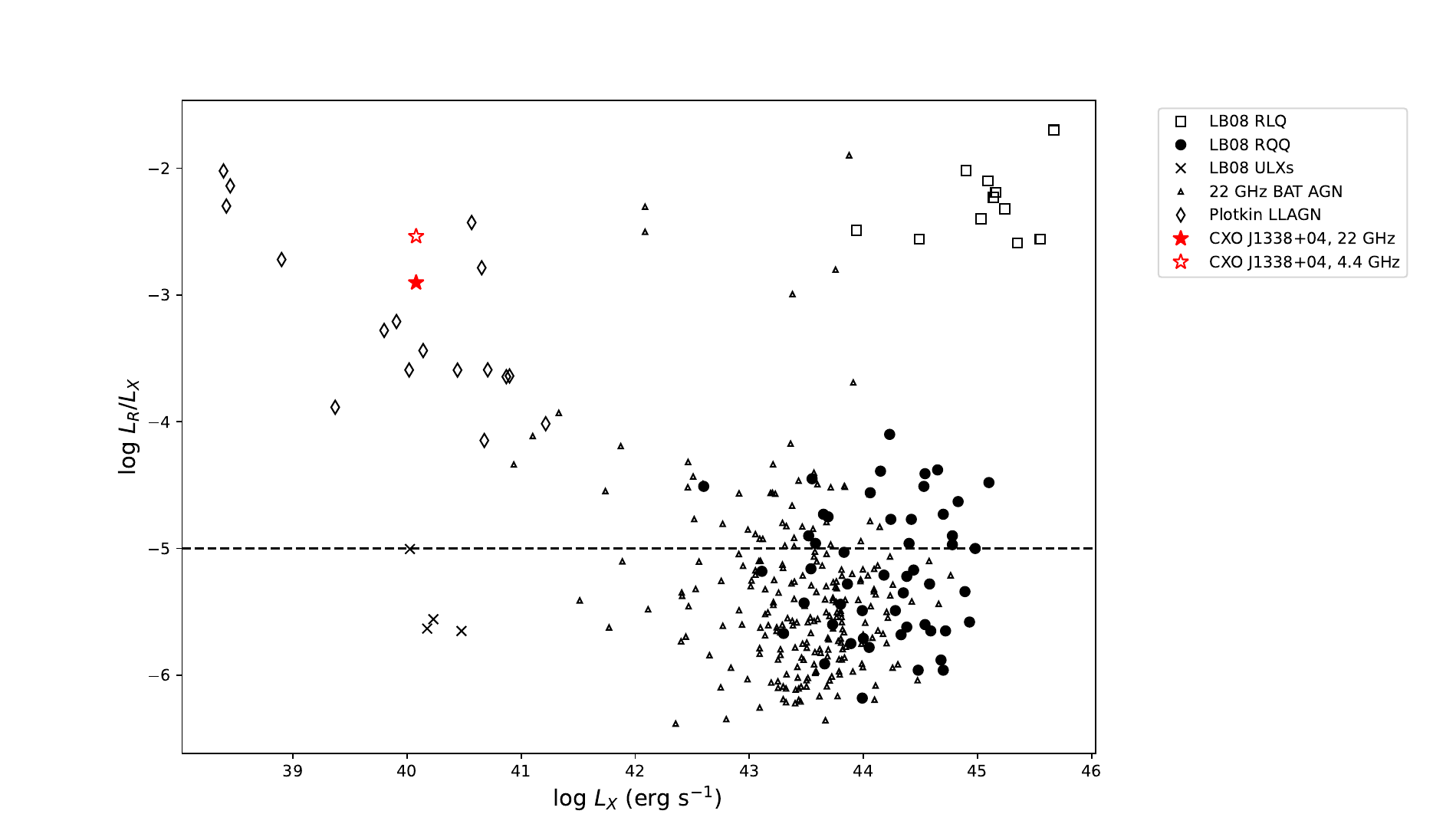}
\caption{Ratio of the radio and X-ray luminosities compared to the X-ray luminosity for the \citet{Laor2008} (LB08) quasar and ULX samples, the BAT AGN Seyfert sample from \citet{Smith2020a}, the LLAGN sample from \citet{Plotkin2012}, and the ULX being studied in this work, CXO~J1338+04. The radio luminosities for the LB08 sample are at 5~GHz for the quasars and a variety of frequencies near 5~GHz for the ULXs. The BAT AGN luminosities are taken from the 1\arcsec core at 22~GHz. We therefore show the values for CXO~J1338+04 measured at 22~GHz by our survey, and at 4.4~GHz by \citet{Mezcua2018}.} 
\label{fig:lrlx}
\end{figure*}

The \citet{Laor2008} comparison also includes a small sample of low-luminosity Seyferts; instead of that sample of 12, which required many corrections of the X-ray and radio fluxes, we include two larger samples: the LLAGN sample from \citet{Plotkin2012}, and the sample of 100 nearby  ($z<0.05$)  Seyferts at 22~GHz from an earlier phase of the same survey in which we detect CXO~J1338+04 here, as well as the radio and X-ray lumionsities of a further $\sim150$ objects that represent the final phase of the survey (and are also radio-quiet Seyferts). The survey will be published fully in an upcoming work (Magno et al., in preparation). The large majority (96\%) of the 22~GHz survey sources are radio-quiet, and are consistent with LLAGN along the fundamental plane of black hole activity \citep{Smith2020a}.

The $0.3-8$~keV X-ray luminosity of CXO~J1338+04 is $L_X = 1.5\times10^{40}$~erg~s$^{-1}$~\citep{Kim2015}. When compared to the 4.4~GHz radio luminosity derived from the flux density of the total jet + core morphology from \citet{Mezcua2018}, this results in a ratio of $L_R/L_X = 1.3\times10^{-3}$. In Figure~\ref{fig:lrlx}, we plot this value along with the objects given in \citet{Laor2008}, the LLAGN of \citet{Plotkin2012}, and the \citet{Smith2020a} high-frequency survey of radio-quiet Seyferts.

Our own observed flux results in a luminosity of $L_\mathrm{22GHz} = 5.2\times10^{37}$~erg~s$^{-1}$, which when compared to the X-ray luminosity yields $L_R/L_X = 4.3\times10^{-3}$. 

We note that the samples compared in this plot were taken from a diverse array of literature sources, and were not all obtained at the same radio or X-ray frequencies or energies. Canonically, the threshold between radio-loud and radio-quiet AGN was proposed by \citet{Terashima2003} as $R_X \equiv \nu L_{\nu,5\mathrm{GHz}} / L_{X,2-10\mathrm{keV}} \sim -4.5$. \citet{Laor2008} also use the radio luminosity at 5~GHz, but in X-rays use the integrated $0.2-20$keV  obtained with the conversion factor $C_\nu=6.25$ (see preceding discussion in this section). Therefore, the X-ray luminosities of objects in Figure\ref{fig:lrlx} may shift left or right by a factor of $\sim5$, or 0.5~dex. The effect of using different bands on the $L_R/L_X$ ratio can be visually estimated by the spread between our 22~GHz data point and the corresponding 4.4~GHz data point from \citet{Mezcua2018}; although their data point is taken with VLBI at significantly smaller spatial scales, it remains above ours by about 0.3~dex. All of the \citet{Smith2020a} points are taken at 22~GHz, and remain mostly consistent in $L_R/L_X$ with the radio-quiet quasars at 5~GHz from \citet{Laor2008} (despite being at lower X-ray luminosities, as expected for Seyferts as compared to luminous quasars). We have used the \emph{Swift-XRT} 2-10~keV X-ray luminosities (rather than the ultra-hard \emph{Swift-BAT} luminosities) when computing the $L_X$ of the \citet{Smith2020b} sample, for maximum consistency with the other samples. It is therefore unlikely that the distinct populations on the plot are due to the observations being taken at different bands or energies; and in any case, CXO~1338+04 retains its relative position to the other populations at both radio frequencies and resolutions presented.

\section{Discussion}
\label{sec:discussion}

When combined with archival observations, our high-frequency data point extends the radio SED significantly. The best-fitting spectral index of $\alpha=-0.66\pm0.02$ is consistent with the canonical value for steep spectrum radio emission from a synchrotron jet \citep[e.g., ][]{Krolik1999}. Because the resolved VLBI observations of \citet{Mezcua2018} indicated a possible core and jet-lobe morphology at much smaller scales, it is reasonable to assume that the larger-scale unresolved radio emission probed by the archival data and our new observation shown in Figure~\ref{fig:seds} is due to this same jet. 
 
Radio-loud and radio-quiet samples show a clear dichotomy in their $L_R/L_X$ ratios. The ratio between the radio and X-ray luminosity of CXO~J1338+04 is consistent with it being a low-mass analog of a radio-loud AGN or quasar, as shown in Figure~\ref{fig:lrlx}. The same conclusion was reached by \citet{Mezcua2018} based on the 4.4~GHz flux of only the western component of their resolved EVN image, which they believe is the location of the core. We have added a point to Figure~\ref{fig:lrlx} that shows the ratio for the entire integrated 4.4~GHz flux from their observation, as well as for our unresolved 22~GHz observation. Both points are far above the radio-quiet quasars from \citet{Laor2008} and from the 95\% radio-quiet BAT AGN sample from \citet{Smith2020a}. In fact, the $L_R/L_X$ ratio is most consistent with radio-loud quasars and the LLAGN sample from \citet{Plotkin2012}, which was chosen to be analogous to the ``low-hard" accretion state of X-ray binaries. The origin of radio emission in radio-loud objects is reasonably well-established as a classical jet, and the LLAGN population is considered most likely to produce jets in accordance with the disk-jet coupling model due to its position on the same fundamental plane as low-hard X-ray binaries and the fact that most LLAGN are radio-loud \citep{Merloni2003,Falcke2004,Terashima2003}, this observation further supports the interpretation of the radio emission from CXO~J1338+04 as a jet launched by an accreting intermediate-mass black hole. In this case, it joins the recently discovered jet candidate found in the dwarf elliptical galaxy SDSS~J090613.77+561015.2 \citep{Yang2023}, which is likely to have a black hole mass of $3.6\times10^5 M_\odot$ \citep{Baldassare2016}. 

It is also apparent from Figure~\ref{fig:lrlx} that CXO~J1338+04 is much more radio-loud than the other ULXs in the \citet{Laor2008} sample. These objects include NGC~5408~X-1, which is likely to be a stellar-mass object accreting at super-Eddington rates and driving a wind nebula \citep{Pinto2016,Luangtip2021}; Homberg~II~X-1, which is unlikely to be $>100M_\odot$ and accreting at slightly sub-Eddington \citep{Goad2006,Ambrosi2022}; M~82~X1, whose nature is uncertain but may be a few$\times100M_\odot$ black hole \citep{Pasham2014,Mondal2022}; and NGC~7424~ULX2, which is located in a young OB association but does exhibit a steep radio spectrum like our source \citep{Soria2006}. In short, most of these objects are either stellar mass objects, or at least are well below the mass range established for CXO~J1338+04 by previous investigations: $10^{3.5} M_\odot < M_\mathrm{BH} < 10^{6.3} M_\odot$ from \citet{Mezcua2018} and $10^5 M_\odot$ by \citet{Kim2020}.  In fact, the majority of ULXs may be stellar mass objects accreting at or above the Eddington rate \citep[e.g., ][]{Berghea2008},  so if CXO~J1338+04 is a bona-fide intermediate-mass black hole producing a real analog of a radio-loud quasar jet, we might expect its $L_R/L_X$ ratio to exceed that of other ULXs, as we observe here. 

\section{Conclusions}
\label{sec:conclusions}

We have presented a new high-frequency 22~GHz radio observation of the IMBH candidate CXO~J133815.6+043255, an ultra-luminous X-ray source in the outskirts of the Seyfert galaxy NGC~5252. We find the following:

\begin{itemize}
    \item The 22~GHz flux density is $S_\nu = 0.53\pm0.02$~mJy. When combined with archival observations, this results in a broadband SED well-fit by a power law with a spectral index of $\alpha = -0.66\pm0.02$ between 1.4 and 22~GHz, consistent with steep radio spectra from synchrotron jet emission.
    \item The $L_R/L_X$ ratio definitively occupies the regime of LLAGN and radio-loud quasars, and is not consistent with radio-quiet Seyferts or with ULXs associated with stellar mass objects. 
\end{itemize}

We conclude that the 22~GHz observations support the conclusion that CXO~J133815.6+043255 is an intermediate-mass black hole producing a radio jet, as suggested by the VLBI observations of \citet{Mezcua2018}. That is, CXO~J133815.6+043255 is likely to be a true low-mass analog of radio-loud quasars. 

\begin{acknowledgments}
KLS and MM gratefully acknowledge discussions fostered by the \emph{VLA Sky Survey in the Multiwavelength Spotlight} conference in Socorro, NM in September 2022. AT is supported by NASA grant number 80NSSC22K0741. This research has made use of the CIRADA cutout service at URL cutouts.cirada.ca, operated by the Canadian Initiative for Radio Astronomy Data Analysis (CIRADA). CIRADA is funded by a grant from the Canada Foundation for Innovation 2017 Innovation Fund (Project 35999), as well as by the Provinces of Ontario, British Columbia, Alberta, Manitoba and Quebec, in collaboration with the National Research Council of Canada, the US National Radio Astronomy Observatory and Australia’s Commonwealth Scientific and Industrial Research Organisation. 
\end{acknowledgments}

\bibliography{ulxpaper}{}
\bibliographystyle{aasjournal}



\end{document}